# Crystal chemistry of light metal borohydrides

Yaroslav Filinchuk[*], Dmitry Chernyshov, Vladimir Dmitriev

Swiss-Norwegian Beam Lines (SNBL) at the European Synchrotron Radiation Facility (ESRF), BP-220, 38043 Grenoble, France

**Abstract.** Crystal chemistry of $M(BH_4)_n$, where M is a $2^{nd}$-$4^{th}$ period element, is reviewed. It is shown that except certain cases, the $BH_4$ group has a nearly ideal tetrahedral geometry. Corrections of the experimentally determined H-positions, accounting for the displacement of the electron cloud relative to an average nuclear position and for a libration of the $BH_4$ group, are considered. Recent studies of structural evolution with temperature and pressure are reviewed. Some borohydrides involving less electropositive metals (e.g. Mg and Zn) reveal porous structures and dense interpenetrated frameworks, thus resembling metal-organic frameworks (MOFs). Analysis of phase transitions, and the related changes of the coordination geometries for M atoms and $BH_4$ groups, suggests that the directional $BH_4$…M interaction is at the origin of the structural complexity of borohydrides. The ways to influence their stability by chemical modification are discussed.

## Introduction

Borohydrides, also called tetrahydroborates, are largely ionic compounds with a general formula $M(BH_4)_n$, consisting of metal cations $M^{n+}$ and borohydride anions $BH_4^-$. Due to a high weight percent of hydrogen, they are considered as prospective hydrogen storage materials. Indeed, some borohydrides desorb a large quantity of hydrogen (up to 20.8 wt %), although the decompositon temperatures are usually high. The search for better hydrogen storage materials, with denser structures and lower binding energies, has been hampered by a lack of basic knowledge about their structural properties. However, the last few years have seen a remarkably rapid evolution of this field, and a large number of experimental and theoretical studies have recently been carried out, as summarized in several very recent reviews on crystal structures (Filinchuk *et al.*, 2008b), physical properties (Sundqvist and Andersson, 2009) and phase relations (Sundqvist, 2009) of borohydrides at different P-T conditions. Different fundamental aspects of the use of light metal hydrides, including borohydrides, for the storage and production of hydrogen have been comprehensively reviewed by Grochala and Edwards, 2004 and more recently by Orimo *et al.*, 2007 and Züttel *et al.*, 2007. Thermodymanic properties of metal borohydrides with respect to their hydrogen storage applications were briefly reviewed by Soloveichik, 2007. This chapter, which is an update of the recent review by Filinchuk *et al.*, 2008b, has a special focus on crystal structures and crystal chemistry of borohydrides.

Almost all borohydrides are crystalline solids already at room temperature. This gives an advantage of using diffraction methods to study their structure. Diffraction provides an immense ammount of information not only about the structure, but also is more and more frequently used for screening and characterization of new substances, reaction products and intermediates. Detailed diffraction study of a promising material at various temperatures and pressures allows to uncover new polymorphs, follow their structural evolution and therefore get a clue to understand (and maybe even alter) the thermodymanic stability. This review will be mostly focused on crystal structures of known metal borohydride phases, also aiming to give a feeling about how much modern diffraction techniques can go beyond a simple structure characterization.

We will review here the experimentally observed structures of light borohydrides $M(BH_4)_n$, where by "light" we assume M to be an element of the $2^{nd}$ to the early $4^{th}$ period. Since the most of the theoretically predicted structures remain hypothetical, this chapter is limited to the experimentally observed ones. The known metal borohydride phases and their crystallographic characteristics are summarized in Table 1. The Table, and this short review do not pretend to be exhaustive, so only one (normally the first) work aiming to determine the crystal strucutre using a given diffraction technique is mentioned here.

---

[*] Corresponding author (e-mail: Yaroslav.Filinchuk@esrf.fr)

**Table 1**. Known phases of light metal borohydrides and their crystallographic characteristics.

| Compound | Space group | $a$ (Å) | $b$ (Å) | $c$ (Å) | $\beta$ (°) | Method[1] | Conditions[2] | Ref. |
|---|---|---|---|---|---|---|---|---|
| $LiBH_4$ | $Pnma$ | 7.17858(4) | 4.43686(2) | 6.80321(4) | | XRPD | | Soulié et al., 2002 |
| | | 7.1213(2) | 4.4060(1) | 6.6744(2) | | NPD | 3.5K | Hartman et al., 2007 |
| | | 7.141(5) | 4.431(3) | 6.748(4) | | SCXRD | 225K | Filinchuk et al., 2008a |
| | $P6_3mc$ | 4.27631(5) | | 6.94844(8) | | XRPD | 408K | Soulié et al., 2002 |
| | | 4.2667(2) | | 6.9223(8) | | NPD | 400K | Hartman et al., 2007 |
| | | 4.3228(10) | | 7.037(1) | | SCXRD | 535K | Filinchuk et al., 2008a |
| | $Ama2$ | 6.4494(9) | 5.307(1) | 5.2919(9) | | XRPD | 2.4 GPa | Filinchuk et al., 2008c |
| | $Fm\text{-}3m$ | 5.109(2) | | | | XRPD | 18.1 GPa | Filinchuk et al., 2008c |
| $NaBH_4$ | $Fm\text{-}3m$ | 6.148(1) | | | | NPD | | Fischer, Züttel, 2004 |
| | | 6.1308(1) | | | | SCXRD | 200K | Filinchuk, Hagemann, 2008 |
| | $P\text{-}42_1c$ | 4.332(1) | | 5.869(1) | | NPD | 10K | Fischer, Züttel, 2004 |
| | $Pnma$ | 7.297(1) | 4.1166(5) | 5.5692(7) | | XRPD | 11.2 GPa | Filinchuk et al., 2007 |
| $KBH_4$ | $Fm\text{-}3m$ | 6.728(1) | | | | SCXRD | | Luck, Schelter, 1999 |
| | | 6.7306(1) | | | | NPD | | Renaudin et al., 2004 |
| $NH_4BH_4$ | $Fm\text{-}3m$ | 6.978 | | | | XRPD | | Karkamkar et al., 2009 |
| | $P4_2/nmc$ | 4.7004(2) | | 6.5979(3) | | NPD | 1.5K | Renaudin et al., 2004 |
| $Be(BH_4)_2$ | $I4_1cd$ | 13.62(1) | | 9.10(1) | | SCXRD | | Marynick, Lipscomb, 1972 |
| $Mg(BH_4)_2$ | $P6_1$ | 10.3182(1) | | 36.9983(5) | | XRPD+NPD | | Černý et al., 2007 |
| | $P6_122$ | 10.354(1) | | 37.055(4) | | SCXRD | 100K | Filinchuk et al., 2009c |
| | $Fddd$ | 37.072(1) | 18.6476(6) | 10.9123(3) | | XRPD | | Her et al., 2007 |
| $Ca(BH_4)_2$ | $Fddd$ | 8.791(1) | 13.137(1) | 7.500(1) | | XRPD | | Miwa et al., 2006 |
| | $F2dd$ | 8.7759(3) | 13.0234(4) | 7.4132(2) | | XRPD | 91K | Filinchuk et al., 2009a |
| | $P4_2/m$ | 6.9468(1) | | 4.3661(1) | | XRPD+NPD | 480K | Buchter et al., 2008 |
| | $P\text{-}4$ or $P4_2nm$ | 6.9189(1) | | 4.3471(1) | | XRPD | 305K | Filinchuk et al., 2009a |
| | $I\text{-}42d$ | 5.8446(3) | | 13.228(1) | | XRPD | 495K | Filinchuk et al., 2009a |
| | $Pbca$ | 13.0584(8) | 8.3881(4) | 7.5107(4) | | XRPD+NPD | 300K | Buchter et al., 2009 |
| $Mn(BH_4)_2$ | $P3_112$ | 10.435(1) | | 10.835(2) | | XRPD | | Černý et al., 2009a |
| $Al(BH_4)_3$ | $C2/c$ | 21.917(4) | 5.986(1) | 21.787(4) | 111.90(3) | SCXRD | 150K | Aldridge et al., 1997 |
| | $Pna2_1$ | 18.021(3) | 6.138(2) | 6.199(1) | | SCXRD | 195K | Aldridge et al., 1997 |
| $LiK(BH_4)_2$ | $Pnma$ | 7.91337(5) | 4.49067(3) | 13.8440(1) | | XRPD | | Nickels et al., 2008 |
| $LiSc(BH_4)_4$ | $P\text{-}42c$ | 6.07593(6) | | 12.0338(1) | | XRPD | | Hagemann et al., 2008 |
| $NaSc(BH_4)_4$ | $Cmcm$ | 8.170(2) | 11.875(3) | 9.018(2) | | XRPD | | Černý et al., 2009b |
| $LiZn_2(BH_4)_5$ | $Cmca$ | 8.6244(3) | 17.8970(8) | 15.4114(8) | | XRPD | | Ravnsbæk et al., 2009a |
| $NaZn_2(BH_4)_5$ | $P2_1/c$ | 9.397(2) | 16.635(3) | 9.136(2) | 112.66(2) | XRPD | | Ravnsbæk et al., 2009a |
| $NaZn(BH_4)_3$ | $P2_1/c$ | 8.2714(16) | 4.5240(7) | 18.757(3) | 101.69(1) | XRPD | | Ravnsbæk et al., 2009a |

[1] SCXRD – single-crystal X-ray or synchrotron diffraction, XRPD – X-ray or synchrotron powder diffraction, NPD – neutron powder diffraction.
[2] Ambient pressure and temperature is assumed if not otherwise specified.

# Crystal structures

## LiBH$_4$

Four LiBH$_4$ phases are known: two at ambient and two at high pressure. The first structural study of the ambient pressure polymorphs was made using synchrotron powder diffraction by Soulié et al., 2002. The low-temperature structure has *Pnma* symmetry. It transforms into a hexagonal wurtzite-like high-temperature phase at ~380K.

Since the originally proposed *P*6$_3$*mc* symmetry of the high-temperature phase has been questioned by theoreticians, it has been revisited by neutron powder diffraction on the triply isotopically substituted $^7$Li$^{11}$BD$_4$ (Hartman et al., 2007) and by synchrotron diffraction on single crystals and powder (Filinchuk, Chernyshov, 2007; Filinchuk et al., 2008a). These recent studies confirm the *P*6$_3$*mc* space group symmetry for the high-temperature phase, and reveal large and anisotropic displacements for hydrogen atoms. The observed atomic displacement ellipsoids (Fig. 1b) may indicate, in addition to the thermal vibrations, an orientational disorder (hindered rotations) of the BH$_4$ group. Large libration amplitudes in the hexagonal phase agree with an apparent shortening of the B-H bonds to ~1.07 Å. An entropy contribution from the disorder is considered as a factor stabilizing the hexagonal structure (Filinchuk, Chernyshov, 2007; Filinchuk et al., 2008a).

Single-crystal investigation of the *Pnma* phase at 225K unambiguously shows that the BH$_4$ group has a nearly ideal tetrahedral geometry, in contrast to strongly distorted tetrahedra reported in earlier synchrotron diffraction studies (Soulié et al., 2002; Züttel et al., 2003). The observed H-B-H angles, 108.8(9)-109.9(7)°, are very close to the ideal tetrahedral angle of 109.5°, and the B-H bond lengths show a narrow spread from 1.104(11) to 1.131(15) Å. Neutron diffraction at 3.5K also shows that the BH$_4$ group is very close to the undistorted tetrahedron: B-H distances are 1.208(3)-1.225(6) Å, H-B-H angles – 107.2(3)-111.7(4)°.

The two ambient pressure phases have similar structures (Figs. 1a and 1b), where Li atoms and BH$_4$ groups are tetrahedrally coordinated. In the orthorhombic phase at 225K, Li-B distances are 2.37-2.57 Å. In the hexagonal phase at 535K, there are three short B-Li contacts of 2.55 Å in the basal plane and one long of 3.00 Å along the *c* axis.

At room temperature and a pressure of 1.2-10 GPa, LiBH$_4$ forms a new phase with pseudo-tetragonal *Ama*2 structure (Filinchuk et al., 2008c). It can be considered as an orthorhombically distorted antistructure of PtS, where Li atoms are tetrahedrally coordinated by BH$_4$ groups (Li-B distances of 2.35-2.66 Å) and the BH$_4$ group has a totally new nearly square-planar coordination by four Li atoms (Fig. 1c). Above 10 GPa another LiBH$_4$ phase forms (Filinchuk et al., 2008c). It is isostructural to the cubic NaBH$_4$ (see below); Li atoms and BH$_4$ groups are octahedrally coordinated (at 18 GPa, the Li-B distances are 2.56 Å). In the *Ama*2 phase, the geometry of the BH$_4^-$ anions was restrained to an ideal tetrahedron with the B-H bond lengths centered at ~1.17 Å, while in the cubic phase at 18.1 GPa the refined B-H distance is 1.08 Å.

The borohydride groups are connected to Li atoms mostly via the tetrahedral edges. However, in the hexagonal phase and in the case of the shortest Li-B contact (2.37 Å) in the *Pnma* phase, the borohydride groups are connected to Li atoms via the tetrahedral faces.

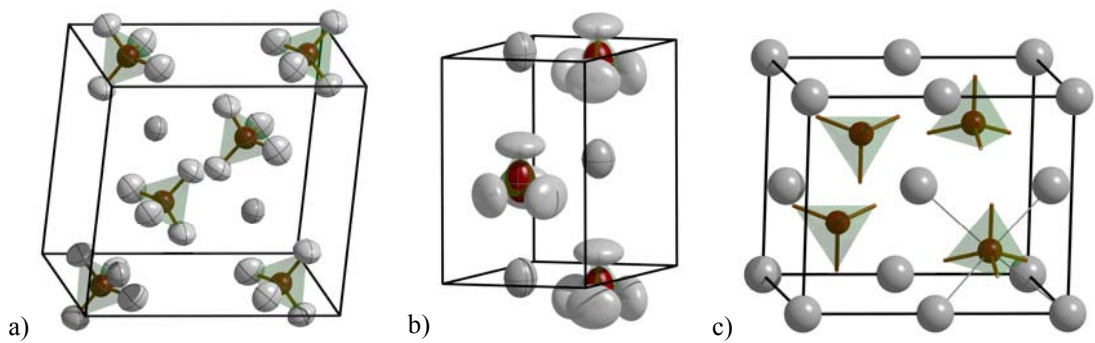

a)                  b)                  c)

**Fig. 1.** Crystal structure of LiBH$_4$ phases: *Pnma* (**a**), *P*6$_3$*mc* (**b**) and *Ama*2 (**c**). The cubic *Fm*-3*m* structure (not shown here) is identical to the one presented in Fig. 2a. Anisotropic displacement ellipsoids determined from synchrotron diffraction on single crystals (Filinchuk et al., 2008a) are shown for the ambient pressure phases.

## NaBH$_4$

Under ambient conditions NaBH$_4$ has a cubic structure, isomorphous to NaCl. From the diffraction data alone, this structure can be desribed in either *Fm*-3*m* or *F*-43*m* space groups. The two models differ only by the absence of the inversion centre in *F*-43*m*. The centrosymmetric model is intrinsically disordered, showing two orientations of the BH$_4^-$ group (Fig. 2a), while the non-centrosymmetric model allows the BH$_4^-$ anion to be fully or partly ordered. A peak of heat capacity at ~190K (Johnston, Hallett, 1953) suggests an order-disorder transition. The estimated entalpy of the transition is consistent with a transformation from the fully orientationally disordered high-tempreture cubic phase to an ordered low-temperature tetragonal phase. Therefore, the cubic phase has to be assumed to have the *Fm*-3*m* space group symmetry (Stockmayer, Stephenson, 1953). Crystal-chemical analysis shows that the shortest, repulsive H…H interactions in the cubic phase favour the disorder: a number of the shortest H…H distances between $BH_4^-$ anions in the disordered *Fm*-3*m* structure is twice smaller than in a hypothetical ordered *F*-43*m* structure (Stockmayer, Stephenson, 1953). Thus, among the two possible models, the disordered *Fm*-3*m* and the ordered *F*-43*m*, only the first agrees both with crystallographic and thermodynamic data, as well as with the crystal-chemical considerations.

The cubic phase was characterized at 293K by neutron diffraction on $NaBD_4$ powder (Fischer, Züttel, 2004) and at 200K by synchrotron diffraction on a $NaBH_4$ single crystal (Filinchuk, Hagemann, 2008). The fully disordered *F*-43*m* model assumed in the first study is equivalent to the conventional *Fm*-3*m* model used in the second one. The ideal tetrahedral geometry of the borohydride group is defined by the site symmetry for the boron atom. The B-D distance at room temperature is ~1.17 Å, and at 200K the apparent B-H bond length is 1.09(2) Å. The latter, being corrected for 0.10 Å increment which takes into account the displacement of the electron cloud relative to an average nuclear position of the H-atom (Filinchuk, Hagemann, 2008), makes 1.19 Å.

On cooling below ~190K or upon compression to ~6 GPa at room temperature a phase with closely related ordered tetragonal structure appears (Kumar, Cornelius, 2005; Sundqvist, Andersson, 2006). Thermal conductivity study allowed to map the P-T boundary between the cubic and tetragonal phases at low temperature (Sundqvist, Andersson, 2006). The tetragonal structure was reported in the *P*-42$_1$*c* space group (Fischer, Züttel, 2004), but as was pointed out (Filinchuk *et al.*, 2007), it deviates insignificantly from the higher symmetry, and thus can also be described in the space group *P*4$_2$/*nmc*. Comparing to the cubic phase, the $BH_4$ groups in the tetragonal phase are ordered in two different orientations (see Figs. 2a and 2b). As a result, a number of the shortest H...H contacts in the tetragonal phase is reduced by one third comparing to the disordered cubic phase, and the network of the repulsive H…H contacts changes from the isotropic three-dimensional (3D) to the two-dimensional (2D) one, oriented in the *ab* plane. Consequently, the *c*/*a* ratio changes discontinuously upon the cubic-to-tetragonal (c-to-t) transition at ~186K from 1 to 0.964 (Filinchuk *et al.*, 2009b); this change is more due to the contraction of the *c*-axis than to the expansion in the basal plane. We also found a small but ubrupt volume drop of 0.40% upon the c-to-t transition at ambient pressure. Both discontinuities confirm that the transition is of the first order.

Neutron powder diffraction at 10K shows that the $BD_4$ group in the tetragonal phase is nearly ideally tetrahedral and the B-D distance is 1.22 Å (Fischer, Züttel, 2004). Both the cubic and the tetragonal phases are strongly textured when loaded in diamond anvil cells (Filinchuk *et al.*, 2007; Chernyshov *et al.*, 2008).

Above ~9 GPa yet another phase was detected by diffraction and Raman spectroscopy, but both experimental and theoretical efforts originally failed to identify its structure (Kumar, Cornelius, 2005; Araújo *et al.*, 2005). Later its structure has been solved from synchrotron powder diffraction data measured at 11.2 GPa, and it was shown to be of $BaSO_4$ type (Fig. 2c). For the successful solution of the structure, it was essential to model a texture, including one parameter in the global optimization (Filinchuk *et al.*, 2007). The structure was solved with most *a*-axes approximately aligned with the compression direction. Hydrogen atoms were located as a part of a semi-rigid $BH_4$ group, with B-H distances of ~1.17 Å. A small volume drop of ~1% was found upon the pressure-induced transition from the tetragonal to the orthorhombic phase (Filinchuk *et al.*, 2007).

In all three structures, Na atoms and $BH_4$ groups are octahedrally coordinated. In the cubic phase at 200K Na-B distances are 3.065 Å, in the tetragonal phase at 180K – 2.976-3.091 Å (Filinchuk *et al.*, 2009b), and in the orthorhombic phase at room temperature and 11.2 GPa – 2.763-2.849 Å.

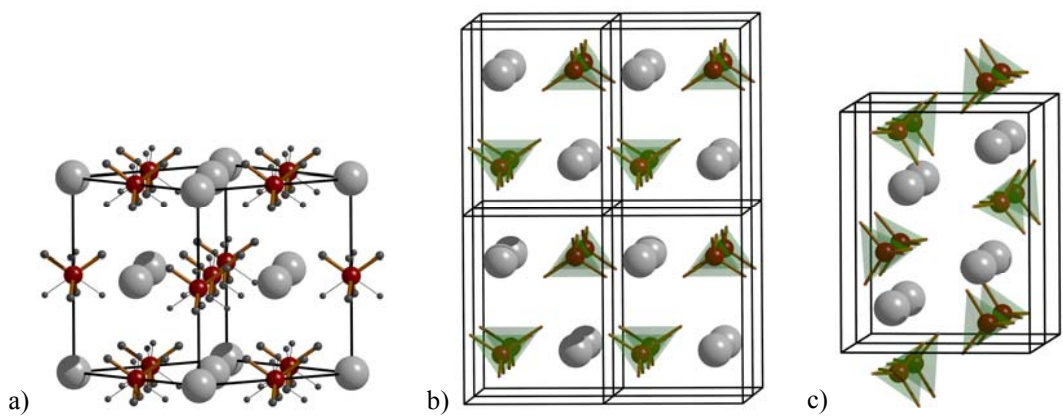

**Fig. 2.** Crystal structure of $NaBH_4$ phases: *Fm*-3*m* (**a**), *P*-42$_1$*c* or *P*4$_2$/*nmc* (**b**) and *Pnma* (**c**). Two orientations of the disordered $BH_4$ group in the cubic phase are shown by bold and thin lines.

## KBH$_4$

This substance shows a similar behaviour to NaBH$_4$. Its cubic phase has been studied at room temperature by X-ray diffraction on single crystals (Luck, Schelter, 1999) and by neutron powder diffraction (Renaudin *et al*., 2004): in both cases it was described in the space group *Fm*-3*m*. The cubic phase transforms into a tetragonal one at 65-70K. Its structure is equivalent to the tetragonal NaBH$_4$, but it was refined in the higher symmetry space group, *P*4$_2$/*nmc*. Temperature dependence of the unit cell parameters for both phases has been reported by Renaudin *et al*., 2004. As the cation's size in the cubic MBH$_4$ increases, as M goes from Na to K and then to Rb and Cs, the unit cell expands and the shortest H…H distances between the neighbouring BH$_4$ anions increase. In this way a weaker H…H repulsion in the heavier MBH$_4$ reduces the temperature of the c-to-t transition. The K-B distance in the cubic KBH$_4$ is 3.364 Å.

X-ray diffraction on KBH$_4$ single crystal at 293K reveals an apparent (uncorrected, for an analysis of corrections see below) B-H distance 1.09(1) Å (Luck, Schelter, 1999). The B-D distance in the tetragonal KBD$_4$, determined by neutron powder diffraction at 1.5K, is 1.205(3) Å, and in the cubic structure at 295K it is 1.196(3) Å (Renaudin *et al*., 2004). It was shown that the B-D distance in the cubic MBH$_4$ at 295K increases as M goes from Na to Cs. Besides the ionic size effect, this may be related to the increased anharmonicity in lighter MBH$_4$ compared to the heavier MBD$_4$ (Renaudin *et al*., 2004), which is presumably related to an increased libration amplitudes for the BH$_4$ group. The latter would lead to an apparent (illusory) shortening of the B-H bonds.

## NH$_4$BH$_4$

Ammonium borohydride contains ca. 24 wt % of hydrogen, and releases more than 20 wt % hydrogen in three steps at temperatures below 433K (Karkamkar *et al*., 2009). It has the same cubic structure as Na and K borohydrides at ambient conditions, with a slightly bigger cell parameter than of the potassium analogue (see Table 1). The recently revealed crystal structure has not yet been studied in detail; in particular the hydrogen atom positions have not been experimentally determined.

## Be(BH$_4$)$_2$

The only known Be(BH$_4$)$_2$ phase was studies by single crystal X-ray diffraction (Marynick, Lipscomb, 1972). Its tetragonal structure contains helical polymeric chains, where the only independent Be cation is coordinated by two bridging borohydride anions (Be-B distances of 2.00 Å) and one terminal borohydride anion (Be-B distance 1.92 Å). It is striking that the therminal BH$_4$ group is coordinated by the Be atom not via the face but via a tetrahedral edge, just the same way as the bridging BH$_4$ group. This may be due to the repulsive interaction between hydrogen atoms of the neighbouring BH$_4$ groups: the corresponding H...H distances (2.24-2.31 Å) are among the shortest known and Be...BH$_4$ coordination via the tetrahedral face would make these distances even shorter. Be atom has a trigonal-planar environment made by three BH$_4$ groups, and the bridgind BH$_4$ group has a linear Be-B-Be geometry. Thus, the low coordination number for the Be atom leads to a reduced dimensionality (1D) of the polymeric structure, where coordination potential of the BH$_4$ ligands is not fully realized.

The BH$_4$ groups show slightly deformed tetrahedral geometry with H-B-H angles 104-127°. The apparent (uncorrected) B-H distances fall into the 1.08(2)-1.18(3) Å range at 293K.

## Mg(BH$_4$)$_2$

The most stable polymorph, α-Mg(BH$_4$)$_2$, has been reported in *P*6$_1$ space group symmetry by two independent groups (Černý *et al*., 2007; Her *et al*., 2007). In one case (Černý *et al*., 2007) its structure was solved and refined using combination of synchrotron and neutron powder diffraction data. The synchrotron data contributed most of the information, but the use of neutron data allowed to determinate the orientation of the rigid BH$_4$ tetrahedra. In the other case (Her *et al*., 2007), the structure was solved from synchrotron powder data alone, however, only positions of Mg and B atoms were reliably determined. Geometry optimization of the experimentally determined hexagonal structure, by using the density functional theory (DFT), suggested that it has a pseudo or maybe even true *P*6$_1$22 symmetry (Dai *et al*., 2008). The structure has been recently revised in the space group *P*6$_1$22, using 100K single-crystal diffraction data (Filinchuk *et al*., 2009c). The revision supports the suggestion made in the DFT study. Analysis of the published *P*6$_1$ models shows that the location of the H-atoms from powder data posed the main problem for the identification of the correct symmetry: wrongly determined orientations of some BH$_4$ groups hampered a successful detection of the true *P*6$_1$22 symmetry by automatic algorithms.

Pure α-phase transforms into the β-phase above 490K, the latter is quenched (metastable) on cooling, showing an anomalous temperature dependence of the cell parameters (Filinchuk *et al*., 2009c). Sharp diffraction peaks of β-Mg(BH$_4$)$_2$ can be described by a relatively simple *Immm* structure. However, all peaks were modeled, assuming an antisite disorder, in the 8 times bigger supercell with *Fddd* symmetry (Her *et al*., 2007).

An intriguing aspect of the Mg(BH$_4$)$_2$ structures is their complexity (Fig. 3). The hexagonal structure contains three symmetry independent Mg$^{2+}$ cations and six symmetry independent BH$_4^-$ anions. Mg atoms bridged by the BH$_4$ groups (Mg-Mg 4.6-4.9 Å) form a framework with a novel 3D topology. It contains a variety of 5-membered (-Mg-BH$_4$-)$_5$ rings that dominate in number over 6- and 8-membered ones, and in this way the structure resembles an amorphous state. The orthorhombic structure contains two symmetry independent Mg$^{2+}$ cations and five symmetry independent BH$_4^-$ anions. Mg atoms bridged by the BH$_4$ groups (Mg-Mg 4.6-4.7 Å) also form a 3D framework, however the latter does not contain 5-membered rings, but only 4- and 6-membered ones.

On the local level, both phases have the same principles of structural organization. Each Mg atom is surrounded by four BH$_4$ tetrahedra in a strongly deformed tetrahedral environment, while each BH$_4$ is nearly linearly coordinated by two Mg cations via the opposite tetrahedral edges. However, the Mg-H$_2$BH$_2$-Mg fragments are not strictly linear – in the α-phase the Mg-B-Mg angles range from 148 to 170°. At ambient conditions, the Mg-B distances in the hexagonal phase vary within the nearly the same interval, 2.39-2.45 (Filinchuk *et al.*, 2009c) as in the orthorhombic phase, 2.34-2.47 Å (Her *et al.*, 2007). With respect to the metal-hydrogen bond lengths and coordination geometry for the metal atoms, Mg(BH$_4$)$_2$ takes an intermediate position between Be(BH$_4$)$_2$ (flat trigonal coordination by three BH$_4$ anions) and LiBH$_4$ (nearly ideal tetrahedral coordination by four BH$_4$ anions). Thus, the ionic metal-to-BH$_4$ size ratio may be one of the reasons of the high structural complexity of Mg(BH$_4$)$_2$. Strong distortion of inequivalent Mg polyhedra provides a sufficient flexibility to meet all the crystal-chemical (i.e. energy) requirements: minimization of the repulsive Mg…Mg and H…H interactions, meeting coordination preferences for the BH$_4$ groups etc.

Accurate single-crystal diffraction data revealed that the H-B-H angles in the Mg-H$_2$B fragments are in average by 2° more open than for the non-coordinated ones. The DFT model shows the same trend. A detectable distortion of the tetrahedral BH$_4$ groups can be attributed to the strong polarizing effect of the Mg$^{2+}$ cation or to a partially covalent Mg-BH$_4$ interaction. Unexpectedly, α-Mg(BH$_4$)$_2$ contains an unoccupied void, accounting for 6.4% of space in the structure (Filinchuk *et al.*, 2009c). It is large enough (37 Å$^3$) to accommodate a small molecule, such as H$_2$O (see Fig. 3). The high-temperature β-phase is less dense by ~3%, but contains no unoccupied voids.

Unusual crystal chemistry and the high structural complexity of Mg(BH$_4$)$_2$ is unprecedented for this class of compounds, and all known attempts to predict Mg(BH$_4$)$_2$ structures theoretically were so far unsuccessful.

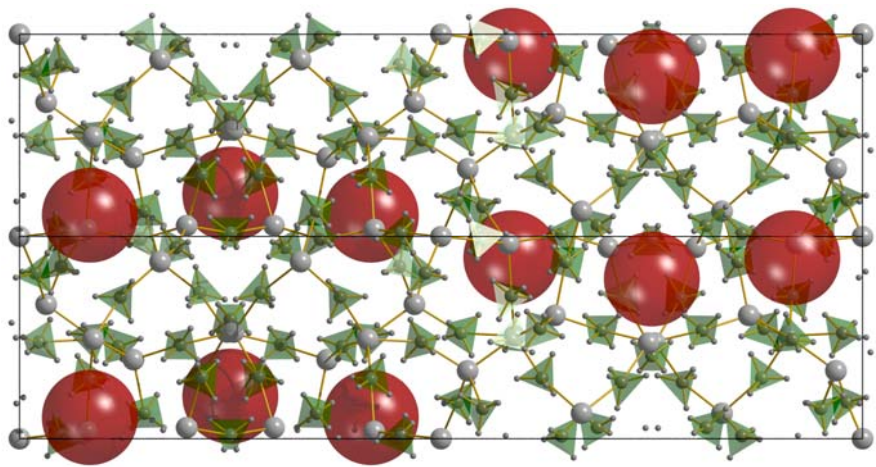

**Fig. 3.** Crystal structure of α-Mg(BH$_4$)$_2$, where the unoccupied voids are shown as large spheres.

## Ca(BH$_4$)$_2$

Riktor *et al.*, 2007 reported on existence of several Ca(BH$_4$)$_2$ phases at different temperatures. For the α-phase, obtained by desolvation of the tetrahydrofurane complex, Miwa *et al.*, 2006 reported an *Fddd* structure. However, analysis of the high-quality synchrotron powder diffraction data revealed that the actual symmetry of this orthorhombic structure is lower (Filinchuk *et al.*, 2009a). Two noncentrosymmetric subgroups of *Fddd*, namely *F2dd* and *Fdd2*, provide a much better fit to the experimental data than the centrosymmetric *Fddd* model, and the *F2dd* structure has some advantage over the *Fdd2* one. The arrangement of the BH$_4$ groups in the noncentrosymmetric structure forms a simpler ordered pattern than in the *Fddd* model. The structure and symmetry of the orthorhombic phase is maintained from low temperatures up to 495K. However, as the temperature increases, the cell parameters *a* and *c* continuously approach each other, and at ~495K a second order transition into a tetragonal *I-42d* α'-phase is taking place (Filinchuk *et al.*, 2009a). The two structures show a group-subgroup relation (index 2). It is important that BH$_4$ groups are fully ordered in the *F2dd* and *I-42d* structures (Figs. 4a and 4b), while tetragonal supergroups of *Fddd* impose a disorder of the borohydride group in a hypothetical high-temperature *I*4$_1$/*amd* structure.

Above 450K, the closely related *F2dd* and *I-42d* phases transform into a completely different β-phase. Determination of its space group symmetry is ambiguous, and its structure was described in two similar models, giving the lowest DFT energies, namely in the space groups *P*4$_2$/*m* (Buchter *et al.*, 2008) and *P-4* (Filinchuk *et al.*, 2009a), see Fig. 4c. It is likely that the BH$_4$ group in β-Ca(BH$_4$)$_2$ is intrinsically disordered, similar to the high-temperature phase of LiBH$_4$ (Filinchuk *et al.*, 2008a), and according to the systematic absences of the diffraction peaks (Filinchuk *et al.*, 2009a) the true symmetry of the disordered structure is *P4₂nm*. The high-temperature phase is by 4-6% (depending on temperature) denser than the *F2dd* and *I-42d* ones, and it is stable on cooling to room and even to liquid nitrogen temperatures (Filinchuk *et al*., 2009a). However, Fichtner *et al*. (2008) found that this phase is meta-stable at room temperature, as it slowly transforms back into the orthorhombic phase. Another phase, called γ-Ca(BH$_4$)$_2$, obtained by wet chemical synthesis, has an orthorhombic structure (Buchter *et al*., 2009), see Fig. 4d. It is metastable at all temperatures and irreversibly transforms into the β-phase at ~590K. At ambient conditions the γ-phase is slightly (~1% or less) denser than the β-phase.

Crystal structures of all four Ca(BH$_4$)$_2$ phases (α, α', β and γ) contain calcium cations nearly octahedrally coordinated by six borohydride anions. Ca-B distances are all within the similar range: 2.82-2.97 Å in the *F2dd* α-phase at 91K, 2.94-2.98 Å in the *I-42d* α'-phase at 495K, 2.92-2.94 Å in the *P-4* β-phase at 305K and 2.86-2.97 Å in the *Pbca* γ-phase. For the first three structures listed above, the refined geometry of the BH$_4^−$ anions was restrained to an ideal tetrahedral configuration with the B-H bond lengths centered at ~1.17 Å (Filinchuk *et al*., 2009a), while for the γ-phase the hydrogen atoms were refined independently yielding a distorted tetrahedral geometry with B-H bond length 1.06(3)-1.33(3) Å, at variance with the theoretical calculations suggesting ideal tetrahedral geometry of the BH$_4$ group (Buchter *et al*., 2009).

In the *F2dd* and *I-42d* structures of the α and α'-phases Ca atoms form a close-packed diamond-like frameworks where the BH$_4$ groups have a T-shaped coordination. On the contrary, a trigonal-planar coordination of the BH$_4$ group by Ca atoms is observed in the β and γ-phases.

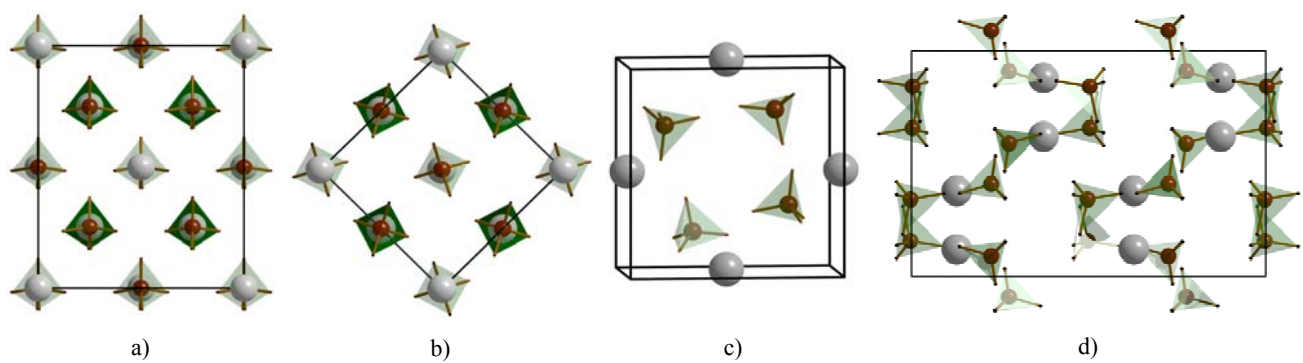

a)          b)          c)          d)

**Fig. 4.** Crystal structure of Ca(BH$_4$)$_2$ polymorphs: α-phase in *F2dd* (**a**), α'-phase in *I-42d* (after the *F2dd* → *I-42d* transition) (**b**), β-phase in *P-4* or *P4₂/m* (the true structure is likely disordered with *P4₂nm* symmetry) (**c**) γ-phase in *Pbca* (**d**).

## Mn(BH$_4$)$_2$

The first crystal structure of a 3*d*-metal borohydride has been determined only recently for Mn(BH$_4$)$_2$ by Černý *et al*., 2009a. The compound forms at ambient conditions in ball-milled mixtures of alkali metal borohydrides and MnCl$_2$, crystallizes with the space group symmetry *P3₁12* and is stable from 90 up to 450K, where the compound melts. Thermal expansion of Mn(BH$_4$)$_2$ between 90 and 400K is highly anisotropic and very non-uniform.

Similar to magnesium borohydride, both independent metal atoms in Mn(BH$_4$)$_2$ are surrounded by four BH$_4$ tetrahedra in deformed tetrahedral coordinations with Mn-B distances ranging 2.39-2.52 Å. Each BH$_4$ group is approximately linearly coordinated by two Mn atoms (Mn-B-Mn angles vary from 147.4 to 170.0(2)°) via the tetrahedral edges. A single B-H distance of 1.13(2) Å was refined for rigid BH$_4$ tetrahedra.

The structure of Mn(BH$_4$)$_2$ shows an interesting similarity to α-Mg(BH$_4$)$_2$: the two structures are made of similar layers. The layers are stacked along the *c*-axis, and rotated by 120° by the 3$_1$ axis in Mn(BH$_4$)$_2$ and by 60° by the 6$_1$ axis in α-Mg(BH$_4$)$_2$. Three identical layers are stacked along one unit cell vector *c* in Mn(BH$_4$)$_2$, six layers are stacked in α-Mg(BH$_4$)$_2$. In Mn(BH$_4$)$_2$ the layers are connected directly while in α-Mg(BH$_4$)$_2$ the layers are connected via a thin intercalated layer which contains one Mg atom per layer per cell. Similar to α-Mg(BH$_4$)$_2$ (Filinchuk *et al*., 2009c), the structure of Mn(BH$_4$)$_2$ is not densely packed and contains isolated voids with the estimated volume of 21 Å$^3$ each, which occupy in total 6% of the space.

## Al(BH$_4$)$_3$

At ambient conditions, this substance is liquid. However, two crystalline phases are known at low temperatures, with a transition temperature in the 180-195K range. Their structures have been studied by single-crystal X-ray diffraction (Aldridge *et al*., 1997). Each phase is made up of discrete Al(BH$_4$)$_3$ units, where Al has a trigonal-planar environment with Al-B distances varying in a very narrow interval, 2.10(2)-2.14(2) Å. All BH$_4$ groups are coordinated by metal atoms via the tetrahedral edges. Locally, the Al(BH$_4$)$_3$ structures resemble the one for Be(BH$_4$)$_2$. However, the shortest H…H distances between neighbouring borohydride anions are longer (an average is over 2.5 Å) than those in Be(BH$_4$)$_2$ (2.24-2.31 Å).

The BH$_4$ anions appear somewhat distorted, partly due to experimental inaccuracies; however these distortions correlate with the results of the electron diffraction studies in the gas phase and *ab initio* calculations (Aldridge *et al.*, 1997). For example, the B-H distances are expected to be longer for the H-atoms coordinated by Al. Indeed, the B-H distances in the B-H…Al bridges are 1.12(3)-1.14(4) Å versus 0.99(4)-1.01(4) Å for the terminal B-H bonds. Note, that the experimental distances are much shorter than the theoretical ones (respectively, 1.27 and 1.19 Å), since they have to be corrected for the displacement of the electron cloud relative to an average nuclear position of the H-atom and for the libration shortening of the B-H bond (Filinchuk *et al.*, 2008a).

## LiK(BH$_4$)$_2$

Very recently a number of mixed-cations borohydrides have been obtained and structurally characterized by synchrotron powder diffraction. These compounds are fully stoichiometric, as they do not show statistical occupation of one site by two different cations.

In the first bimetallic borohydride, LiK(BH$_4$)$_2$ (Nickels *et al.*, 2008), Li atom is tetrahedrally surrounded by BH$_4$ groups at Li-B distances 2.51-2.61 Å. An increase of the coordination number for K atom from six in KBH$_4$ to seven in LiK(BH$_4$)$_2$ increases also the K-B distances from 3.36 Å to 3.40-3.48 Å. Two independent BH$_4$ anions are coordinated via the tetrahedral edges. One of them has a typical octahedral coordination (Li$_2$K$_4$), similar to that in the cubic MBH$_4$ phases, while the other shows a new type of coordination – square-pyramidal (Li$_2$K$_2$ in the base + K in the apical position). The distortion of the BH$_4$ anions observed in LiK(BH$_4$)$_2$ is rather related to experimental inaccuracies than to an influence of the polarizing cations suggested by Nickels *et al.*, 2008, for the discussion on the BH$_4$ geometry see below.

## MSc(BH$_4$)$_4$ (M = Li, Na)

LiSc(BH$_4$)$_4$ (Hagemann *et al.*, 2008) can be described as a complex containing [Sc(BH$_4$)$_4$]$^-$ anion (Borisov, Makhaev, 1988), similar to other structurally characterized anion borohydride complexes, such as (Ph$_4$P)$_2$[Mg(BH$_4$)$_4$] (Makhaev *et al.*, 2004). Vibrational spectroscopy and DFT calculations indicate the presence of discrete [Sc(BH$_4$)$_4$]$^-$ ions. Sc atoms have a distorted tetrahedral coordination by BH$_4$ groups (Sc-B distances of 2.28 Å), the four BH$_4$ groups are oriented with a tilted plane of three hydrogen atoms directed to the central Sc ion, resulting in a 8 + 4 coordination. The Li ions are found to be disordered along the *z* axis of the tetragonal cell, thus the coordination of the Li atom is not clearly defined. The borohydride group in LiSc(BH$_4$)$_4$ was modelled as a rigid body with a common refined B-H distance 1.08 Å.

The structure of NaSc(BH$_4$)$_4$ (Černý *et al.*, 2009b) consists of the similar isolated [Sc(BH$_4$)$_4$]$^-$ anions located inside slightly deformed trigonal prisms of Na atoms. Sc atom coordinates the BH$_4$ groups via the faces, yielding a 12-fold coordination by H atoms. Na is surrounded by six BH$_4$ tetrahedra in quite regular octahedral coordination with a (6+12)-fold coordination by H atoms. The packing of Na$^+$ cations and [Sc(BH$_4$)$_4$]$^-$ anions in NaSc(BH$_4$)$_4$ is a deformation variant of the hexagonal NiAs structure type.

## MZn$_2$(BH$_4$)$_5$ (M = Li, Na)

LiZn$_2$(BH$_4$)$_5$ represent a novel type of structure (Ravnsbæk *et al.*, 2009a) which has no distinct analogues among other known inorganic compounds. The structure of NaZn$_2$(BH$_4$)$_5$ (Ravnsbæk *et al.*, 2009a) was identified and refined as a monoclinically distorted derivative of the Li-containing analogue.

Two independent Zn atoms in MZn$_2$(BH$_4$)$_5$, M = Li and Na, have a trigonal nearly planar coordination (coordination number, CN = 3) by three BH$_4$ groups, similar to the Be atoms in Be(BH$_4$)$_2$ (Marynick, Lipscomb, 1972). The Li and Na atoms in these two related structures have a saddle-like coordination (CN = 4), which has not been observed earlier for alkaline atoms in borohydrides. All BH$_4$ groups in MZn$_2$(BH$_4$)$_5$ are linearly coordinated by two metal atoms (the angles in the two structures cover the range 164.5-179.6°), in a similar way as in the Mg(BH$_4$)$_2$ structures. The BH$_4$ groups are coordinated via the two opposite tetrahedral edges, bridging either two Zn atoms or one Zn and one M atom. H-atoms were refined as a part of a semi-rigid BH$_4$ units, with B-H distances close to 1.12 Å and H-B-H angles close to 109.5°.

It is remarkable that MZn$_2$(BH$_4$)$_5$ consists of two identical doubly-interpenetrated three-dimensional (3D) frameworks (Fig. 5a), which implies that there are no covalent bonds between them. This type of structural topology is common for the coordination polymers involving organic ligands, also known as metal-organic frameworks (MOFs), but it is observed for the first time in metal hydrides. This suggests directionality and some covalent character of the metal-BH$_4$ interaction. Indeed, the Zn-H bonds in LiZn$_2$(BH$_4$)$_5$ are very short; the refined distances are all below 2 Å, reaching 1.65 Å at the lower limit. Zn-B contacts are also very short, 2.11-2.31 Å. The average Zn-B distance of 2.17 Å in LiZn$_2$(BH$_4$)$_5$ is longer than the corresponding 1.97 Å average in Be(BH$_4$)$_2$ for the trigonally coordinated Be atom, but considerably shorter than in all other metal borohydrides. Considering only the strongly associated Zn atoms and BH$_4$ units, the NaZn$_2$(BH$_4$)$_5$ compounds contain isolated [Zn$_2$(BH$_4$)$_5$]$^-$ anions (Fig. 5b), counter-balanced by cations M$^+$, similar to the isolated [Sc(BH$_4$)$_4$]$^-$ units in MSc(BH$_4$)$_4$.

It was recently found that the three-dimensional framework in Mg(BH$_4$)$_2$ contains empty voids, large enough to accommodate small molecules like H$_2$O (Filinchuk *et al.*, 2009c). MZn$_2$(BH$_4$)$_5$ structures, although not porous, reveal another case of strong and directional metal-BH$_4$ bonding defining the structural architecture.

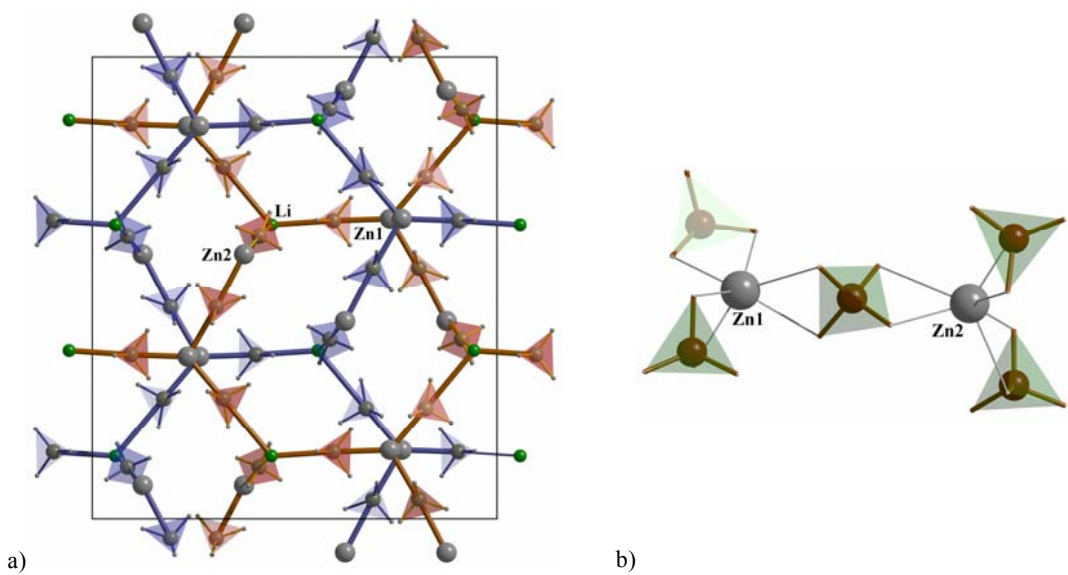

**Fig. 5.** Crystal structure of LiZn$_2$(BH$_4$)$_5$. The doubly interpenetrated three-dimensional framework is highlighted in blue and red (**a**). Trigonally coordinated Zn atoms and the BH$_4$ groups are associated into the isolated [Zn$_2$(BH$_4$)$_5$]$^-$ anions (**b**).

### NaZn(BH$_4$)$_3$

This phase shows a new type of structure (Ravnsbæk *et al*., 2009a), containing a 3D framework made of the metal atoms and the BH$_4$ groups. The Zn atoms in NaZn(BH$_4$)$_3$ have a distorted tetrahedral coordination with CN = 4, while the saddle-like coordination of the Na atom by BH$_4$ groups is similar to the one in MZn$_2$(BH$_4$)$_5$. Comparing to MZn$_2$(BH$_4$)$_5$, the change of the coordination number from three to four for Zn leads to much longer Zn-B distances, varying from 2.43 to 3.16 Å. The average Zn-B distance of 2.74 Å in NaZn(BH$_4$)$_3$ is much longer than the 2.42 Å average of the Mg-B distances in the hexagonal Mg(BH$_4$)$_2$, where the metal atoms also have a tetrahedral coordination by the BH$_4$ groups (Filinchuk *et al*., 2009c). Interestingly, while one o the BH$_4$ groups in NaZn(BH$_4$)$_3$ exhibits a nearly linear coordination by two metal atoms, the other two show a trigonal-planar coordination.

## Structural evolution under non-ambient conditions

### Diffraction studies of the stability regions and structure evolution under external stimuli

The examples shown above illustrate that powder diffraction provides accurate information about structure of borohydrides. In some of the best examples, its accuracy is comparable to the one of the single crystal diffraction. However, a profound use of this technique is to study structure evolution under variable pressure or temperature. Bulk modulus and coefficients of thermal expansion, together with information on the symmetry and structure, provide the basis for validation of theoretical models. A study of structure evolution with external stimuli thus appears to be complementing and verifying the theoretical calculations.

Most light borohydrides show rich polymorphism indicating a complex profile of the potential energy. LiBH$_4$ serves as an illustration of the complex behaviour in a seemingly simple system (Filinchuk *et al*., 2008a). While the high-temperature hexagonal phase shows a uniform and isotropic temperature expansion, the low-temperature *Pnma* phase reveals non-linear and highly anisotropic response (Fig. 6). The cell dimension *b* continuously contracts on heating from 300K to the transition temperature. Such thermal expansion reflects an anharmonicity of the potential of the crystal binding, which should be taken into account in theoretical models. It is remarkable that the thermal expansion is also anomalous at low temperatures: the parameter *a* deviates from linear dependence below 200K, shows a minimum at ~150K, and then increases on cooling. This observation can be related to a thermodynamic bistability involving the high-pressure *Ama*2 phase. Indeed, according to thermal conductivity measurements at high pressures, the free energy of LiBH$_4$ below 180K and ambient pressure should have two minima (Talyzin *et al*., 2007). It is remarkable that the thermal expansion in the *a*-direction deviates from the linear behaviour in the same temperature range. A similar bistability and even more pronounced anomalous cell expansion is revealed for the two phases of Mg(BH$_4$)$_2$ (Filinchuk *et al*., 2009c), that may be related to the evolution of the free energy profile.

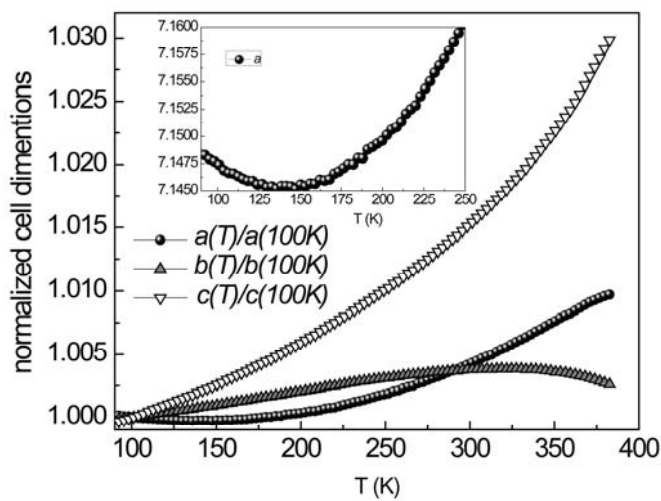

**Fig. 6.** Cell dimensions of the *Pnma* phase of LiBH$_4$ as a function of temperature (Filinchuk *et al.*, 2008a), scaled to the experimental values at 100K. The low-temperature behaviour of the *a* parameter is shown in the inset.

The need to consider high-pressure phases while explaining the temperature behaviour of LiBH$_4$ calls for a combined pressure-temperature (P-T) study of the corresponding phase diagram. *In situ* synchrotron diffraction serves as a best probe to map the P-T diagram (Fig. 7), identify the phases and follow their structural evolution (Dmitriev *et al.*, 2008). Such a diagram, with all the supporting information, allows evaluating fundamental thermodynamic and structural properties of LiBH$_4$ and may guide a rational chemical modification of LiBH$_4$. A pressure behaviour of the transition temperature allows to estimate an entropy change on the corresponding transition (Pistorius, 1974), and thus to draw a conclusion about its mechanism. An analysis of symmetry changes and structural deformations, followed by a group theoretical analysis, yields a unified picture of the phase transformations in LiBH$_4$. The results of such study are shortly presented below for LiBH$_4$. A similar investigation of the P-T diagram, followed by a group-theoretical and crystal-chemical analysis has been very recently made for another light hydride, ammonia borane (Filinchuk *et al.*, 2009d).

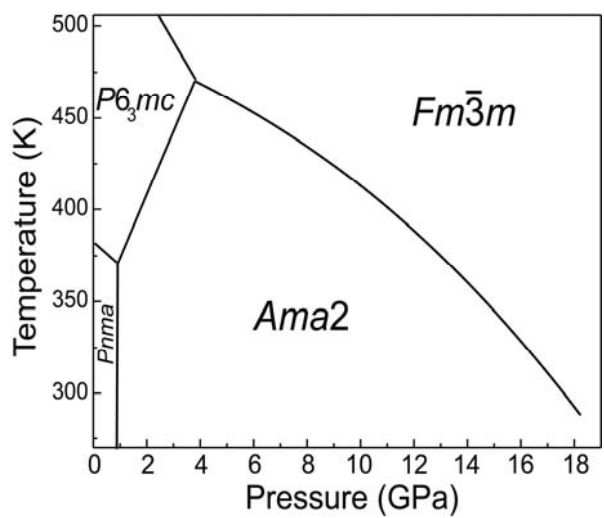

**Fig. 7.** Pressure-temperature phase diagram of LiBH$_4$ from synchrotron diffraction experiments (Dmitriev *et al.*, 2008).

**Phenomenological and crystal-chemical analysis of the mechanisms of the phase transitions**

An existence of cation-anion layers in all four LiBH$_4$ phases is suggested from the phenomenological analysis of mechanisms of phase transitions (Dmitriev *et al.*, 2008). This conclusion is not trivial from purely geometrical point of view, but it can find a rational explanation considering the experimentally determined structures. A clear evidence for the existence of cation-anion layers is found in the hexagonal phase, where BH$_4$ tetrahedron has three short Li-B contacts in the *ab* plane and a long one along the *c* axis. In the *Pnma* phase, these layers are corrugated (Fig. 8a) and the structure is less anisotropic. In the high-pressure phases the existence of cation-anion layers is less obvious. However, the layers where Li and BH$_4$ groups are associated by means of the shorter Li-B contacts can be identified in the (011) plane of the *Ama*2 structure. Due to the high symmetry of the cubic structure different hypothetical layers can be identified there.

However, only one type of layers is consistent both with geometrical considerations and with the phenomenological model: these are cation-anion layers situated in the (111) plane, very similar to those found in the hexagonal phase.

Clearly, the formation of layers in the LiBH$_4$ structures is not determined by coordination polyhedra for Li and BH$_4$ groups, since corresponding coordination numbers and geometries vary with pressure and temperature. However, directional coordination of the BH$_4$ group by Li atoms indicates geometrical and possibly also electronic preferences of BH$_4$…M interaction. Interaction of non-spherical BH$_4$ anions with spherical Li cations results in cation-anions layers, which determine the mechanisms of transitions between polymorphic structures. Directional interaction of the tetrahedral BH$_4$ with the spherical metal atoms explains the relative complexity of LiBH$_4$ structures and of the P-T phase diagram in comparison with NaCl, where both the cation and the anion are spherical.

It has been shown (Dmitriev *et al.*, 2008) that the order parameter in LiBH$_4$ can be parameterized as a shift of layers formed by Li and BH$_4$, together with in-layer deformations. The temperature dependence of Li-B distances can be, therefore, taken as a probe of inter- and intralayer deformations. Indeed, it was shown (Hagemann *et al.*, 2009) that the layers defined in the *bc* plane of the orthorhombic phase (Fig. 8a) become more flat as the temperature is increased: the difference between atomic coordinates for Li and B atoms decreases with temperature. Also, the distance between the layers significantly increases with temperature (Fig. 8b). Thus, the thermal evolution of the crystal structure agrees with the phase diagram and description of the order parameters suggested by the Landau theory.

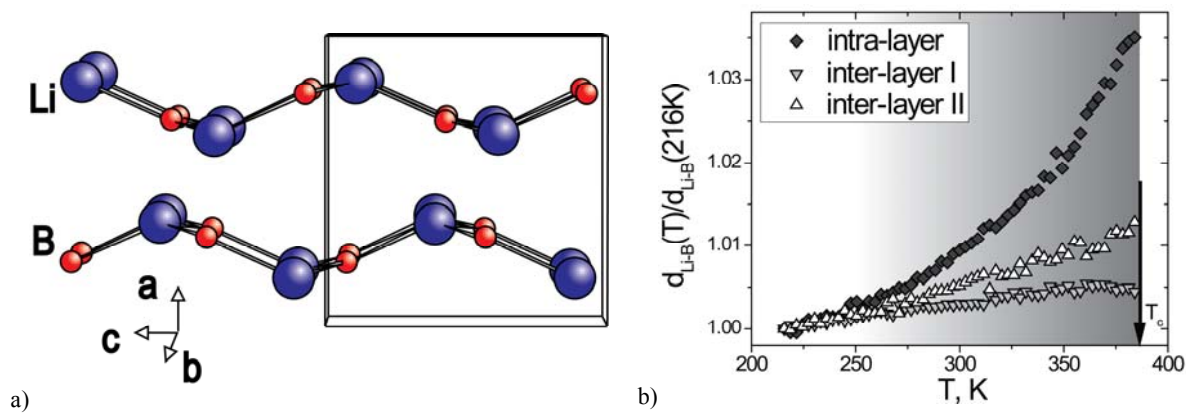

**Fig. 8.** Li-BH$_4$ layers (H atoms omitted) in the orthorhombic phase of LiBH$_4$ (**a**) and the temperature evolution of Li-B distances within and between the layers (**b**). The gradient background highlights an increase of lattice anharmonicity, the arrow indicates the temperature of the transition into the hexagonal phase.

Another application of combined crystal-chemical and phenomenological analysis of polymorphic transitions would be to reveal destabilization of borohydrides upon phase transitions, achieved via formation of short H…H distances and deformation of tetrahedral BH$_4$ anions. A sign of such destabilization was observed in the *Ama*2 phase of LiBH$_4$ (Filinchuk *et al.*, 2008c). A similar destabilization in LiBH$_4$ or similar systems may be achieved during reconstructive phase transitions or melting. Indeed, some hydrogen desorption from LiBH$_4$ occurs upon the transition from *Pnma* to *P*6$_3$*mc* phase and on melting (Mauron *et al.*, 2008). It is suggested (Dmitriev *et al.*, 2008) that other phase boundaries in the P-T phase diagram of pure or chemically modified LiBH$_4$ should be addressed with respect to a possible hydrogen desorption. Chemical modification of borohydrides, for example by exchanging a part of BH$_4$ anions by halide anions (Mosegaard *et al.*, 2008; Maekawa *et al.*, 2009; Matsuo *et al.*, 2009; Oguchi *et al.*, 2009; Mosegaard *et al.*, 2009), can be used along with pressure and temperature to investigate their hydrogen desorption properties within the given P-T phase diagram.

# Crystal chemistry

## Geometry of the BH$_4$ group

Distorted geometry of the BH$_4$ group in the *Pnma* phase of LiBH$_4$ (Soulié *et al.*, 2002; Züttel *et al.*, 2003), found from the early synchrotron powder diffraction experiments, posed a question whether strongly polarizing Li$^+$ cation can induce deformations of the covalently bonded and therefore presumably rigid tetrahedral BH$_4^-$ anion.

The recent revision of the LiBH$_4$ structures by single crystal synchrotron diffraction (Filinchuk, Chernyshov, 2007; Filinchuk *et al.*, 2008a) showed that the H-B-H angles are nearly the same as in the ideal tetrahedron (109.5°), while three independent B-H bonds have nearly the same lengths. This showed that the BH$_4$ group is very close to the undistorted tetrahedron and that the spurious distortions are not related to the nature of the interaction between X-rays and the electron density of light hydrogen atoms. Undistorted geometry of the BH$_4$ group in LiBH$_4$ was also established by neutron powder diffraction on isotopically substituted $^7$Li$^{11}$BD$_4$ (Hartman *et al.*, 2007).

Synchrotron powder diffraction data obtained using a two-dimensional (2D) image plate detector also define the $BH_4$ anion as nearly ideally tetrahedral (Filinchuk, Chernyshov, 2007; Filinchuk *et al.*, 2008a). An analysis of the intensity distributions along the diffraction rings suggests that the powder data obtained with a 1D detector, measuring only a small part of Debye rings, suffers from a poor powder average. Intensity integration over the Debye rings recorded with a 2D detector provides much better average over the grains and therefore higher accuracy of the refined structural parameters. It is therefore crucial in the reliable determination of H-atoms on a small-volume powder sample containing relatively large grains. Considerable deviations from the tetrahedral geometry of the $BH_4$ group, as for instance recently reported by Nickels *et al.*, 2008 from synchrotron powder diffraction data, should be explained by an experimental inaccuracy. We should note, however, that in certain borohydride structures the $BH_4$ group can indeed be deformed, provided a good structural reason of this unusual behaviour, such as short H…H contacts, exists (Filinchuk *et al.*, 2008c). Despite it is not observed in $LiBH_4$, an interaction of a strongly polarizing cation with $BH_4^-$ can also lead to deformations, $Al(BH_4)_3$ (Aldridge *et al.*, 1997) and $\alpha$-$Mg(BH_4)_2$ (Filinchuk *et al.*, 2009a) serve as an examples. In any case, such deformation can be reliably detected mainly by diffraction on single crystals or by DFT calculations.

A certain discrepancy exists between the B-H bond lengths determined from X-ray and neutron diffraction experiments. While neutron diffraction at 3.5K gives the refined B-H distances of 1.208(3)-1.225(6) Å (Hartman *et al.*, 2007), the X-ray diffraction at low temperature yields 1.104(11) to 1.131(15) Å (Filinchuk *et al.*, 2008a). A similar difference has been found from accurate diffraction data on other light hydrides. This difference $\delta_{EL}$, illustrated on Fig. 9a, comes from the well-known displacement of the electron cloud (observed by X-ray diffraction) relative to an average nuclear position (seen by diffraction of neutrons). By its nature, this difference is temperature independent.

Furthermore, DFT calculations find the B-H bond lengths still slightly longer than those from neutron diffraction. This second difference $\delta_{LIB}$ (Fig. 9b), originates from a geometric effect, caused by a libration of the $BH_4$ unit, which leads to an underestimation of the experimentally determined distances. An extent of the $BH_4$ libration is not only temperature-dependent (at higher temperature the motion of the $BH_4$ groups is more pronounced) but also, via frequencies of the corresponding phonons, is specific to each crystal structure. The extent of this libration shortening can be estimated from atomic displacement parameters determined from an accurate diffraction experiment.

A combined correction of 0.10 Å has been recently suggested from the analysis of the available structural data for $NaBH_4$ and its hydrate at low temperatures (Filinchuk, Hagemann, 2008). It roughly accounts both for the displacement of the electron cloud on H-atoms and for the libration effect observed in ordered structures at 200-300K. Displacement of H-atoms from B along the B-H bonds brings positions of hydrogen atoms determined from X-ray diffraction in agreement with theoretically predicted values.

The two corrections were evaluated separately from synchrotron diffraction data collected on $LiBH_4$ single crystals. It is shown (Filinchuk *et al.*, 2008a) that the displacement of the electron cloud from the nuclear position of H-atoms contributes ~0.08 Å to the apparent shortening of the B-H bonds. The libration correction in the *Pnma* phase at 225K was estimated to be 0.034 Å, while in the high-temperature phase at 535K it is much bigger, ~0.10 Å.

In is interesting to note here that the libration amplitudes for the $BH_4$ group, along with the related anharmonicities, are larger than for the heavier $BD_4$. Consequently, the cell parameters for $MBH_4$ are slightly larger (usually by less than 1% of the cell volume, Renaudin *et al.*, 2004) than those for $MBD_4$. The different contribution of H and D to the libration correction should also be considered in accurate comparisons of interatomic distances.

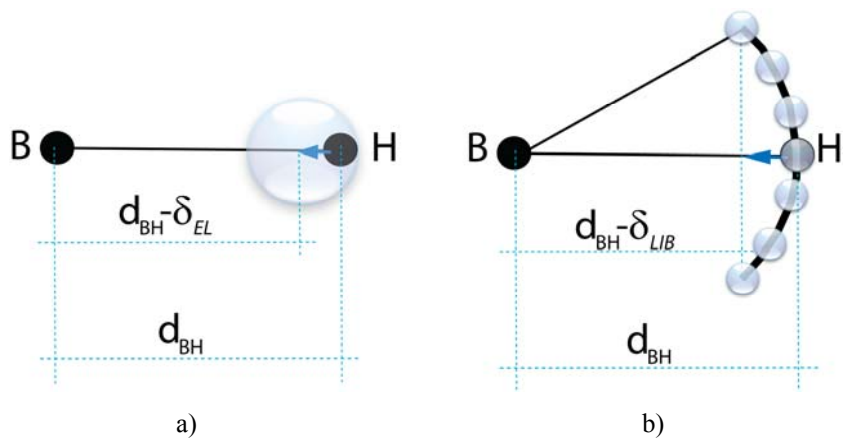

**Fig. 9.** Schematic illustration of systematic differences between B-H distances determined by various methods. A discrepancy $\delta_{EL}$ between the B-H bond lengths determined from X-ray and neutron diffraction (**a**) comes from the displacement of the electron cloud (observed by X-ray diffraction) relative to the average nuclear position (seen by diffraction of neutrons). This difference is temperature independent, $\delta_{EL}$ = 0.08 Å (Filinchuk *et al.*, 2008a). A discrepancy $\delta_{LIB}$ between DFT calculations and the experimentally determined B-H bond lengths originates also from the geometric effect (Fig. 9b) caused by a libration of the $BH_4$ unit. $\delta_{LIB}$ is temperature-dependent, specific to each crystal structure, and dependent on the mass of the H-isotope. Note that for the X-ray diffraction data both corrections have to be applied.

Correction of H-positions determined from X-ray diffraction for systematic errors leads to an increased accuracy of this technique applied to low-Z hydride systems. It also allows to compare directly the results obtained by different diffraction techniques and by theoretical calculations. Making comparisons without applying proper systematic corrections has lead to some odd conclusions (Chater *et al*., 2006; Siegel *et al*., 2007; Yang *et al*., 2007; Wu *et al*., 2008), when neutron powder diffraction giving irregular $BH_4$ groups was considered more accurate than singe crystal X-ray diffraction yielding ideally tetrahedral $BH_4$ groups. When the corrections are applied, we arrive to the opposite conclusion. Indeed, nonoverlapped three-dimensional information uncovered by the scattering of X-rays on single crystals gives an advantage over the neutron powder diffraction, where all scattering is projected on one dimension.

## $BH_4\ldots M$ and $BH_4\ldots BH_4$ contacts

Coordination number (CN) for metal atoms increases with their ionic radii. For Be and Al atoms CN = 3 (trigonal-planar coordination by three $BH_4$ group), for Mg atoms CN = 4 (deformed tetrahedral coordination), for Li atoms at ambient and moderate pressures CN = 4 (tetrahedral coordination), but at higher pressure CN increases to 6 (octahedral coordination). For the largest Na, K and Ca atoms CN = 6 and the coordination is ideal or deformed octahedral. Potassium atom in $LiK(BH_4)_2$ is seven-coordinated, its coordination environment is a capped trigonal prism.

Coordination number for the borohydride anion changes along with $M(BH_4)_n$ shoichiometry. At n = 1, the typical coordination geometries are tetrahedral (ambient pressure phases of $LiBH_4$, $LiK(BH_4)_2$) and octahedral (cubic $LiBH_4$, $NaBH_4$, $KBH_4$ and $LiK(BH_4)_2$). As n increases, low-connected T-shaped ($Ca(BH_4)_2$) and linear ($Be(BH_4)_2$ and $Mg(BH_4)_2$) geometries occur, and in the structures with low dimensionalities ($Be(BH_4)_2$ and $Al(BH_4)_3$) the $BH_4$ group can be coordinated even by only one metal atom. Unusual coordination geometries for the $BH_4$ group are square-planar, found in the *Ama*2 phase of $LiBH_4$, and square-pyramidal in $LiK(BH_4)_2$.

The diversity of coordination geometries for the $BH_4^-$ anion indicates that it behaves as a flexible ligand, adjusting to the requirements (CN) of the metal atom. However, the $BH_4\ldots M$ interaction shows a very directional behaviour. As a rule, the borohydride groups are connected to the metal atoms via the tetrahedral edges. The exceptions are the $MSc(BH_4)_4$ phases and the hexagonal phase of $LiBH_4$, where the $BH_4$ group is coordinated via the tetrahedral faces, and the shortest Li-B contact (2.37 Å) in the *Pnma* phase of $LiBH_4$, where the borohydride group is coordinated by Li atom also via the face. Coordination via vertices was observed only in some $Ca(BH_4)_2$ structures (Filinchuk *et al*., 2009a) and in $KZn(BH_4)Cl_2$ (Ravnsbæk *et al*., 2009b). Directional coordination of the $BH_4$ group by metal atoms clearly indicates non-spherical nature of the $BH_4$ anion, which has geometrical and possibly also electronic preferences in $BH_4\ldots M$ interaction.

The H…H distances between neighbouring $BH_4$ groups exceed 2.2 Å, and in most cases they are even longer. The only exception is the *Ama*2 phase of $LiBH_4$, where at rather low pressure of 2.4 GPa a strikingly short H…H contact between adjacent $BH_4$ anions has been found (1.92 Å in the experimental structure and 1.87 Å in the theoretically optimized model). It was shown that the short H…H interaction distorts the $BH_4$ anion, and this is likely to decrease the activation energy for hydrogen desorption. It was suggested that the internal pressure in the $LiBH_4$ structure may be tuned by a partial substitution of lithium by larger cations, or substitution of some $BH_4$ groups by bigger anions (Filinchuk *et al*., 2008c). The resulting $LiBH_4$-based substance with *Ama*2 structure may show more favourable hydrogen storage properties than pure $LiBH_4$ and may turn out to be useful for hydrogen storage applications.

$BH_4\ldots M$ interaction is also responsible for the formation of doubly-interpenetrated 3D frameworks in $MZn_2(BH_4)_5$ and porous 3D framework in α-$Mg(BH_4)_2$. These structural architectures are common for the coordination polymers involving organic ligands, such as MOFs, but they are observed for the first time in metal hydrides. These observations suggest directionality and some covalent character of the $BH_4\ldots M$ interaction in the borohydrides involving less electropositive metals (e.g. Mg and Zn). This may also explain very complex structures and polymorphism of $Mg(BH_4)_2$.

## Chemical destabilization: mixed-cation and mixed-anion borohydrides

Substitution of metal atoms or $BH_4$ groups by other cations and anions can be considered as a way to alter stability of borohydrides. A few mixed-cations borohydrides have been found recently: $LiK(BH_4)_2$ (Nickels *et al*., 2008) and $MSc(BH_4)_4$ (Hagemann *et al*., 2008; Černý *et al*., 2009b). However, hydrogen desorption temperatures for such compounds appear to be high. For $LiK(BH_4)_2$ it is not lower than for the single cation constituents, but merely an intermediate between the two (Nickels *et al*., 2008).

An apparent linear correlation between Pauling electronegativites and decomposition temperatures for borohydrides has been found (Nakamori *et al*., 2006). It was pointed out (Ravnsbæk *et al*., 2009a) that the significant structural diversity and low decomposition temperatures for the novel series of alkali-zinc borohydrides, such as $MZn_2(BH_4)_5$ and $NaZn(BH_4)_3$, may be assigned to the ability to form more covalent bonds between Zn and $BH_4$ units, as compared to the alkali metal borohydrides. Indeed, the Pauling electronegativity of zinc is higher than those for the alkaline metals and Sc, $Zn(BH_4)_2$ has not yet been isolated in its free form, and its instability may contribute to the lower stability of Zn-based borohydrides. Variation in the ratio between the alkali and transition metal, as well as a use of different metals, thus enables tuning of the properties for hydrogen storage in alkali metal-transition metal-$BH_4$ systems. Therefore, a variety of novel mixed-cation transition-metal based borohydrides may be discovered in the near future.

The mixed-anion derivatives of metal borohydrides show even more favourable hydrogen storage properties than the individual components; this is well illustrated for the LiBH$_4$-LiNH$_2$ system (Meisner *et al*., 2006). Currently, two mixed borohydride-amide phases have been structurally characterized: Li$_4$(BH$_4$)(NH$_2$)$_3$ (Filinchuk *et al*., 2006; Chater *et al*., 2006) and Li$_2$(BH$_4$)(NH$_2$) (Chater *et al*., 2007). Very recently an existence of mixed borohydrides-alanates has been reported: Na$_4$(BH$_4$)(AlH$_4$)$_3$ crystallizes in a primitive cubic 2×2×2 supercell of NaBH$_4$ (Smith *et al*., 2007), Mg(BH$_4$)(AlH$_4$) was identified with yet unknown crystal structure (Zhao *et al*., 2007). Hydrogen desorption properties for the latter class of compounds are not yet reported. Currently, only amide containing mixed-anion derivatives showed some decreasing of the hydrogen evolution temperature comparing to the two hydrogen-containing constituents (Meisner *et al*., 2006). However, we can see that the mixed-anion derivatives show a bigger improvement of hydrogen storage properties than the mixed-cation ones. This may be partly due to the formation of dihydrogen bonds (such as B-H...H-N) between different anions, which are expected to facilitate hydrogen desorption. A similar modification of light metal borohydrides can be achieved by introducing neutral hydrogen-rich molecules, such as ammonia or water. Such examples include Mg(BH$_4$)$_2$·2NH$_3$ (Soloveichik *et al*., 2008) and NaBH$_4$·2H$_2$O (Filinchuk, Hagemann, 2008); the first substance shows very favourable hydrogen storage properties.

Very recently a mixed-anion borohydride has been obtained (Ravnsbæk *et al*., 2009b), containing isolated [Zn(BH$_4$)Cl$_2$]$^-$ anions and octa-coordinated K$^+$ cations, associated with the anions by weak K…Cl and K…BH$_4$ interactions. The Zn$^{2+}$ ion has a trigonal planar coordination by two Cl$^-$ and one BH$_4^-$ ion. KZn(BH$_4$)Cl$_2$ starts to decompose at a significantly lower temperature (383K) than KBH$_4$, hence this study provides new inspiration for design and preparation of novel materials, which may be suitable for hydrogen storage applications.

All the structures of the borohydride derivatives described above are fully ordered and shoichiometric, without statistical occupation of the cation or anion sites. LiBH$_4$ + LiCl is the first system where a gradual replacement of BH$_4$ has been found (Mosegaard *et al*., 2008), showing a partial replacement of the borohydride anions by chloride anions at elevated temperatures (>389K). Comparison of the unit cell volumes for different inorganic salts suggests that the size of anions change according to the sequence I$^-$ > BH$_4^-$ > Br$^-$ > Cl$^-$ (Filinchuk, Hagemann, 2008). This variation gives an efficient tool to tune the unit cell volume and an internal lattice pressure in borohydrides. Larger halide anions, such as iodide, stabilize the high-temperature hexagonal phase of LiBH$_4$ at room temperature, which is interesting as a lithium superionic conductor (Maekawa *et al*., 2009; Matsuo *et al*., 2009; Oguchi *et al*., 2009). Even smaller halide anions, such as chloride, stabilize the hexagonal phase at significantly lower temperatures, the latter depend on the degree of the anion substitution (Mosegaard *et al*., 2009). Heating a LiBH$_4$ + LiCl mixture produces highly chloride-substituted hexagonal lithium borohydride, Li(BH$_4$)$_{1-x}$Cl$_x$, with $x$ ~ 0.42. At higher temperatures the solubility is higher, and the process of the LiCl entrance into LiBH$_4$ is reversible on heating/cooling (Mosegaard *et al*., 2009). The orthorhombic low-temperature LiBH$_4$ phase shows a lower solubility of LiCl than the hexagonal one.

Thus, a partial substitution of the BH$_4$ group by halide anions (Hal) opens the way to modify metal borohydrides and influence their structural stability. It may possibly allow obtaining high-pressure polymorphs of metal borohydrides at ambient conditions (Filinchuk *et al*., 2008a). *In-situ* powder diffraction study of the reaction mixtures at variable temperature should be more widely used to probe different M(BH$_4$)$_n$ + M'Hal combinations.

## Chemical destabilization: substitution in the BH$_4$ group

A note on a possible modification of the BH$_4$ group is warranted. A partial substitution of hydrogen atoms by fluorine atoms has been achieved in alanates, showing that the complex hydride Na$_3$AlH$_{6-x}$F$_x$ is less stable than Na$_3$AlH$_6$ (Brinks *et al*., 2008). Due to the high stability of the covalent B-H bond, a similar chemical modification of the BH$_4$ anion is much more difficult. Some recent reports suggest that BH$_3$OH$^-$ exists in solutions (Ruman *et al*., 2007), and even more interestingly, the first alkoxide-substituted borohydride containing BH$_3$OR$^-$ anion has been isolated in the solid state and structurally characterized (Gainsford *et al*., 2009).

We note, however, that amidoboranes, a new class of high-capacity hydrogen storage materials of the general formula M(BH$_3$NH$_2$)$_n$ (Xiong *et al*., 2007), can be considered as modified borohydrides, where one hydrogen atom in BH$_4^-$ has been exchanged for NH$_2$. Nevertheless, amidoboranes are currently obtained not by modifying borohydrides but from ammonia-borane, which contains the ready-made B-N bond.